\documentclass{edbk} 
  \usepackage{epsfig}

\normallatexbib

\begin{document}

\articletitle[]{Production Of Multiple $^{87}$Rb Condensates And Atom Lasers By Rf
Coupling }

\chaptitlerunninghead{Production Of Multiple $^{87}$Rb Condensates And Atom
  Lasers...}

\author{F. Minardi, C. Fort, P. Maddaloni, and M.
  Inguscio}

\affil{ INFM -- European Laboratory for Non Linear Spectroscopy
  (L.E.N.S.) -- Dipartimento di Fisica dell'Universit\`a di Firenze
  L.go E. Fermi 2, I-50125 Firenze, Italy}

\section{Introduction}

One of the major goals in the study of Bose--Einstein condensation
(BEC) in dilute atomic gases has been the realization and development
of atom lasers. An atom laser may be understood as a source of
coherent matter waves. One can extract coherent matter waves from a
magnetically trapped Bose condensate. Schemes to couple the atomic
beam out of the magnetic trap have been demonstrated
\cite{1ketterle,2bloch,3anderson,4phillips}. In the experiments of
\cite{1ketterle,2bloch} the output coupling is performed by the
application of a radio--frequency (rf) field that induces atomic
transitions to untrapped Zeeman states. The atom laser described in
\cite{3anderson} is based on the Josephson tunnelling of an
optically trapped condensate and in \cite{4phillips} a two--photon
Raman process is described that allows directional output coupling
from a trapped condensate.

Characterizing the output coupler is
necessary to understand the atom laser itself. The literature dealing
with theoretical descriptions of output couplers for Bose--Einstein
condensates has focused both on the use of rf transitions \cite{5,6,7,8} and
Raman processes \cite{9}.

Rf output coupling is based on single-- or multi--
step transitions between trapped and untrapped atomic states. As a
consequence, a rich phenomenology arises that include the observation
of "multiple" condensates corresponding to atoms in different atomic
states. These may display varied dynamical behaviour while in the
trap. Also, both pulsed and continuous output--coupled coherent matter
beams have been observed.  The phenomenology is made even more varied
by the possibility of out--coupling solely under gravity and also of
magnetically pushed out beams. The apparatus operated by the Florence
group \cite{10} offers the possibility to investigate various aspects of
output--coupling achieved by rf transitions of atoms in a magnetically
trapped $^{87}$Rb.

\section{experimental production of the condensate}

We bring a $^{87}$Rb sample to condensation using the now standard
technique of combining laser cooling and trapping in a double
magneto--optical trap (MOT) and evaporative cooling in a
magneto--static trap. Our apparatus had been originally designed for
potassium, as presented by C. Fort in this book \cite{11}. 

Our double MOT set--up consists of two cells connected in the
horizontal plane by a 40~cm long transfer tube with an inner diameter
of 1.1~cm. We maintain a differential pressure between the two cells
in order to optimize conditions in the first cell for rapid loading of
the MOT ($10^{-9}$~Torr) while the pressure in the second cell is
sufficiently low ($10^{-11}$~Torr) to allow for the long trapping
times in the magnetic trap necessary for efficient evaporative
cooling.

Laser light for the MOTs is provided by a cw Ti:sapphire
laser (Coherent model 899-21) pumped with 8~W of light coming from an
Ar$^+$ laser. The total optical power of the Ti:sapphire laser on the
Rb D$_2$ transition at 780~nm is 500~mW. The laser frequency is locked
to the saturated absorption signal obtained in a rubidium vapour cell.
The laser beam is then split into four parts each of which is
frequency and intensity controlled by means of double pass through an
AOM: two beams are red detuned respect to the $F=2 \rightarrow F'=3$
atomic resonance and provide the cooling light for the two MOTs.
Another beam, resonant on the $F=2 \rightarrow F'=3$ cycling
transition, is used both for the transfer of cold atoms from the first
to the second MOT and for resonant absorption imaging in the second
cell. Finally a beam, resonant with the $F=2 \rightarrow F'=2$
transition, optically pumps the atoms in the low field seeking
$F=2$,~$m_F=2$ state immediately before switching on the
magneto--static trap. 5~mW of repumping light for the two MOTs
resonant on the $F=1 \rightarrow F'=2$ transition are provided by a
diode laser (SDL-5401-G1) mounted in external cavity configuration.

In the first MOT, with 150~mW of cooling light split into three
retroreflected beams (2~cm diameter), we can load $10^9$ atoms within
a few seconds.  However, every 300~ms we switch off the trapping
fields of the first MOT and we flash on the "push" beam (1~ms
duration, few mW) in order to accelerate a fraction of atoms through
the transfer tube into the second cell. Permanent magnets placed
around the tube generate an hexapole magnetic field that guides the
atoms during the transfer. In the second cell the atoms are recaptured
by the second MOT which is operated with six independent beams
(diameter=1~cm) each with 10~mW of power. The overall transfer
efficiency between the two MOTs is $\sim30$\%, and after 50 shots we
have typically loaded $1.2 \cdot 10^9$ atoms in the second MOT. The final
part of laser cooling in the second MOT is devoted to maximizing the
density and minimizing the temperature just before loading the
magnetic trap. Firstly the atomic density is increased with 30~ms of
Compressed--MOT \cite{13} and this is followed by 8~ms of optical
molasses to reduce the temperature. Soon after, we optically pump the
atoms into the low-field seeking $F=2$,~$m_F=2$ state by shining the
$\sigma^+$--polarized $F=2 \rightarrow F'=2$ beam for 200~$\mu s$,
together with the repumping light. At this point we switch on the
magneto--static trap in the second cell where we perform evaporative
cooling of the atoms.

\begin{figure}[h,t,b]
\begin{displaymath}
\epsfxsize=7cm
\epsfbox{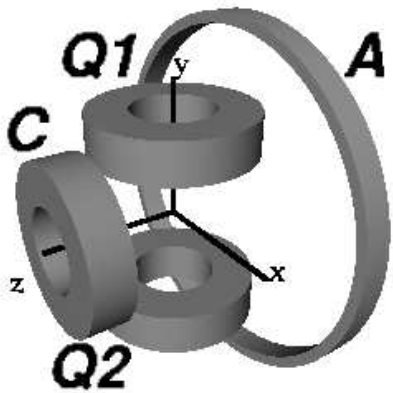}
\end{displaymath}
\caption{4-coils magnetic trap: Q1 and Q2 coils produce the quadrupole
field, the curvature coil (C) provides the axial confinement and the
antibias coil (A) reduces the bias field thus increasing the radial
confinement.}
\label{mag}
\end{figure}

The magneto--static trap is created by passing DC current through
4--coils (see Fig.\ref{mag}), which gives rise to a cigar--shaped
harmonic magnetic potential elongated along the $z$ symmetry axis
(Ioffe--Pritchard type). Our magnetic trap is inspired by the scheme
first introduced in \cite{12}, but is operated with a higher current.
The coils are made from 1/8--inch, water cooled, copper tube.  The
three identical coils consist of 15 windings with diameters ranging
from 3~cm to 6~cm. The fourth coil consists of 6 windings with a
diameter of 12~cm. The coils Q1 and Q2 (Fig.\ref{mag}) generate a
quadrupole field symmetric around the vertical $y$--axis, and in this
direction the measured field gradient is 10~mT/(A $\cdot$ m). These
two coils operated together at low current ($\sim$10~A) also provide
the quadrupole field for operation of the MOT. The {\it curvature} (C)
and {\it antibias} (A) coils produce opposing fields in the $z$
direction. The modulus of the magnetic field during magnetic trapping
has a minimum displaced by 5~mm from the center of the quadrupole
field (toward the curvature coil) and the axial field curvature is
0.46~T/(A$\cdot$m$^2$). The coils are connected in series and fed by a
Hewlett Packard 6681A power supply.  By means of MosFET switches we
can, however, disconnect coils A and C.  The maximum current is 240~A,
corresponding to an axial frequency of $\nu_z$=13 Hz for atoms trapped
in the $F=2$,~$m_F=2$ state. The radial frequency $\nu_r$ can be
adjusted by tuning B$_b$=min(|B|), the bias field at the center of the
trap: $\nu_r$=(2.18/(B$_b$[T])$^{1/2}$)~Hz. With typical operating
values of B$_b$ from 0.14~mT to 0.18~mT, $\nu_r$ ranges from 160~Hz to
180~Hz. In addition, a set of three orthogonal pairs of Helmholtz
coils provide compensation for stray magnetic fields.

The transfer of atoms from the MOT to the magneto-static trap is
complicated by the fact that the MOT (centered at the minimum of the
quadrupole field) and the minimum of the harmonic magnetic trap are
5~mm apart. The transfer of atoms from the MOT to the magnetic trap
consists of a few steps. We first load the atoms in a purely
quadrupole field with a gradient of 0.7~T/m (I=70~A), roughly
corresponding to the ``mode-matching'' condition (magnetic potential
energy equals the kinetic energy) which ensures minimum losses in the
phase--space density. Then we adiabatically increase the gradient to
2.4~T/m by ramping the current to its maximum value I=240~A in 400~ms.
Finally, the quadrupole potential is adiabatically (750~ms)
transformed into the harmonic one by passing the current also through
the {\it curvature} (C) and {\it antibias} (A) coils, hence moving the
atoms 5~mm in the $z$ direction. At the end of this procedure, 30\% of
atoms have been transferred from the MOT into the harmonic magnetic
trap and we start rf forced evaporative cooling with $4 \cdot 10^8$ atoms
at 500~$\mu$K. We estimate the elastic collision rate to be $\gamma
\sim$30~s$^{-1}$ and this, combined with the measured lifetime in the
magnetic trap of 60~s, gives a ratio of "good" to "bad" collisions of
$\sim$1800. This is sufficiently high to perform the evaporative
cooling and reach BEC.

The rf field driving the evaporative cooling is generated by means of
a 10~turn coil of diameter 1--inch placed 3~cm from the center of the
trap in the $x$ direction and fed by a synthesiser (Stanford Research
DS345). The rf field is first ramped for 20~s with a exponential--like
law from 20~MHz to a value which is only 100~kHz above the frequency
 that empties the trap, $\nu_{rf}^{0}=\mu_0$B$_b/2\hbar$. Then a 5~s
linear ramp takes the rf closer to $\nu_{rf}^{0}$: the BEC transition
takes place roughly 5~kHz above $\nu_{rf}^{0}$.

We analyse the atomic cloud using resonant absorption imaging. The
atomic sample is released by switching off the current through the
trapping coils in 1~ms. The cloud then falls freely under gravity and
after a delay of up to 25~ms, we flash a probe beam, resonant with the
$F=2 \rightarrow F'=3$ transition, for 150~$\mu$s and at one tenth of
the saturation intensity.  The shadow cast by the cloud is imaged onto
a CCD array (pixel size=24~$\mu$m) with two lenses, giving a
magnification of 6.  However our resolution is 7~$\mu$m, due to the
diffraction limit of the first lens (f=60~mm, N.A.=0.28).  We process
three images to obtain the two dimensional column density
\~{n}($y,z$)=$\int$ n($x,y,z$)~d$x$. The column density is then fitted
assuming that n($x,y,z$) is the sum of a gaussian distribution
corresponding to the uncondensed fraction and an inverted parabola,
which is solution of the Gross--Pitaevskii equation in the
Thomas--Fermi approximation (condensed fraction). The effect of free
expansion, which is trivial for the gaussian part, is taken into
account also for the condensate as a rescaling of the cloud radii,
according to \cite{14}. The temperature is obtained from the gaussian
widths of the thermal cloud.

We observe the BEC transition at a temperature T$_c$=200~nK with 2$
\cdot 10^5$ atoms, the peak density n$_c$ being 7$ \cdot
10^{19}$~m$^{-3}$. The number of condensed atoms shows fluctuations of
20\% from shot to shot. We may attribute this to thermal fluctuations
of the magnetic trap coils giving rise to fluctuation of the magnetic
field.

\begin{figure}[h,t,b]
\begin{displaymath}
\epsfxsize=10cm
\epsfbox{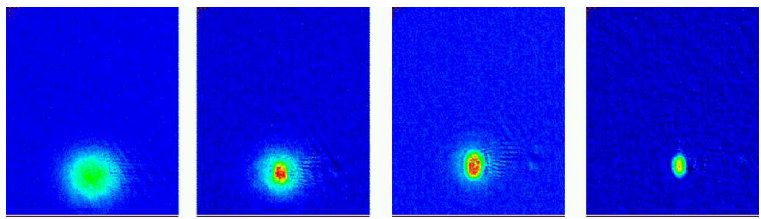}
\end{displaymath}
\caption{Absorption imaging picture of the atomic cloud after a time of
flight of 20~ms. From the left to the right the rf ramp was stopped
respectively at 0.94~MHz, 0.92~MHz, 0.90~MHz and 0.88~MHz. The
symmetric expansion in the first image is typical of a thermal
cloud. By further evaporation the high density peak characteristic of
the condensate emerges, finally giving rise to a pure condensate.}
\label{trans}
\end{figure}

\section{Radio-frequency output coupling}

After producing the condensate in the $F=2$,~$m_F=2$ state, the same
rf field used for evaporation is also employed to coherently transfer
the condensate into different $m_F$ states of the $F=2$ level.
Multistep transitions take place at low magnetic field where the
Zeeman effect is approximately linear. This means that the rf field
couples all the Zeeman sublevels of $F=2$. Two of these, $m_F=2$ and
$m_F=1$, are low--field seeking states and stay trapped. $m_F=0$ is
untrapped and falls freely under gravity, while $m_F=-1$ and $m_F=-2$
are high--field seeking states and are repelled from the trap.
Different regimes may be investigated by changing the duration and
amplitude of the rf field. The absorption imaging with a resonant beam
tuned on the $F=2 \rightarrow F'=3$ transition allows us to detect at
the same time all Zeeman sublevels of the $F=2$ state.

It is worth noting that the spatial extent of the condensate results
in a broadening of the rf resonance. Due to their delocalization
density atoms experience a magnetic field that is non--uniform over
their spatial extent. Our condensate is typically 40~$\mu$m in the axial
direction and 4~$\mu$m in the radial one. The corresponding resonance
broadening is of the order of 1~kHz. This means that rf pulses shorter
than 0.2~ms interact with all the atomic cloud, while for longer
pulses, and sufficiently small amplitudes, only a slice of the
condensate will be in resonance with the rf
field.

\section{pulsed regime}

\begin{figure}[h,t,b]
\begin{displaymath}
\epsfxsize=6cm
\epsfbox{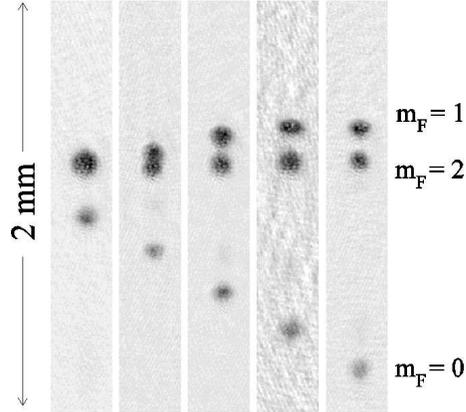}
\end{displaymath}
\caption{Time of flight images of multiple condensates produced by a
  rf pulse. Taken after 15~ms of free fall, the pictures refer to
  different evolution times $\Delta$t in the magnetic trap,
  respectively 2,3,4,5 and 6~ms. The $m_F=2$ condensate is below the
  center of oscillation of the $m_F=1$, opposite to what one expects
  considering the gravitational sagging. This can be explained by the
  presence of a gradient during the switching off of the magnetic
  field}
\label{pulse}
\end{figure}

We investigate the regime of ``pulsed'' coupling characterized by rf
pulses shorter than 0.3~ms. In particular, Fig.\ref{pulse} shows the
effect of a pulse of 10 cycles at $\sim$1.2~MHz (B$_b$=0.17~mT) with
an amplitude B$_{rf}$=7~$\mu$T. After the rf pulse, we leave the
magnetic trap on for a time $\Delta$t and then switch off the trap,
thus allowing the atoms to expand and fall under gravity for 15~ms.
Pictures from the left to the right correspond to trap times after the
rf pulse of $\Delta$t=2,3,4,5 and 6~ms.

Three distinct condensates are visible (Fig.\ref{pulse}): we observe
that one is simply falling freely in the gravitational field and hence
we attribute to the condensate atoms being in the $m_F=0$ state. The
other two condensates initially overlap and then separate. However, we
point out that the pictures are always taken after an expansion in the
gravitational field. The initial position of the condensate in the
trap may be found by applying the equation of motion for free--fall
under gravity.

Leaving the magnetic field on for longer times after the rf pulse
allows us to identify the condensates in different $m_F$ state by
their different center of mass oscillation frequency in the trap.
Considering that the images are taken after a free fall expansion of
t$_{exp}$=15~ms, one can deduce the oscillation amplitude in the trap,
{\it a}, from the observed oscillation amplitude, A, by using the
relation
\begin{equation}
A=a\sqrt{1+\omega_i^2 t_{exp}^2}
\end{equation}
where $\omega_i$ is the oscillation frequency for atoms in the $m_F=i$
level.

From Fig.\ref{pulse} we note that the position of the center of mass
of the condensate in $m_F=2$ is fixed (at the level of resolution)
while the condensate in $m_F=1$ oscillates at the radial frequency of
the corresponding trap potential, with a measured amplitude of {\it
  a}=$8.7 \pm 0.4$~$\mu$m. This can be explained by considering the
different trapping potentials experienced by the two condensates. The
total potential results from the sum of the magnetic and the
gravitational potentials, so that the minima for the two states in the
vertical direction are displaced by $\Delta y$=g/$\omega_r^2$
(``sagging''). With the experimental parameter of Fig.\ref{pulse}
($\omega_r=2 \pi(171 \pm 4$~Hz) for atoms in $F=2$,~$m_F=2$ state)
$\Delta$y equals $8.5 \pm 0.4$~$\mu$m. This is in very good agreement with
the measured center of mass oscillation amplitude. The $m_F=1$ condensate
is produced at rest in the equilibrium position of the $m_F=2$ condensate
and begins to oscillate around its own potential minimum with an
amplitude equal to
$\Delta$y.

\begin{figure}[h,t,b]
\begin{displaymath}
\epsfxsize=8cm
\epsfbox{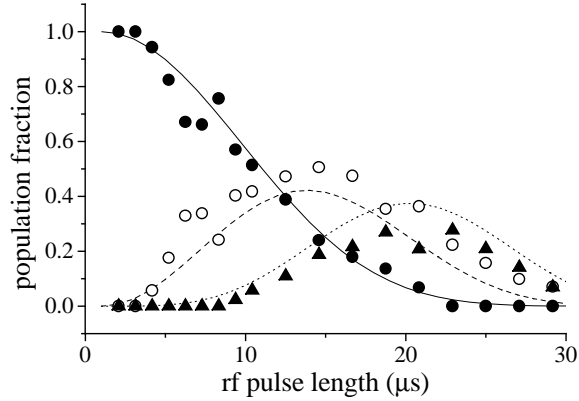}
\end{displaymath}
\caption{Measured population fraction of $m_F=2$ (solid circle), $m_F=1$
(open circle) and $m_F=0$ (triangle) states as a function of rf pulse
duration. The curves correspond to the calculated Rabi oscillations
of the relative population in $m_F=2$ (solid line), $m_F=1$ (dashed line)
and $m_F=0$ (dotted line).}
\label{popo}
\end{figure}

Fig.\ref{pulse} shows the situation where the rf pulse is adjusted to
equally populate the two trapped states, $m_F=1$ and $m_F=2$. In
general, the relative population in different Zeeman sublevels can be
determined by varying the duration of the rf pulse. This is clearly
illustrated in Fig.\ref{popo}, where the relative population of the
$m_F=2,1$ and 0 condensates are shown as a function of the pulse
duration.

The theoretical curves, calculated for a Rabi frequency of 26~kHz
corresponding to the amplitude of our oscillating rf magnetic field
B$_{rf}$=3.6~$\mu$T, are shown together with the experimental data.
The population of each Zeeman state is calculated by solving the set
of the Bloch equation in the presence of an external rf coupling
field.  These results clearly show that we can control in a
reproducible way the relative populations of the multiple condensates.
It is worth noting that the use of a static magnetic trap allows a
straightforward explanation of the phenomenon; similar investigations
recently reported for a time-dependent TOP trap \cite{15} show that
the theory is more complicated in presence of a time varying magnetic
field.

\begin{figure}[h,t,b]
\begin{displaymath}
\epsfxsize=2cm
\epsfbox{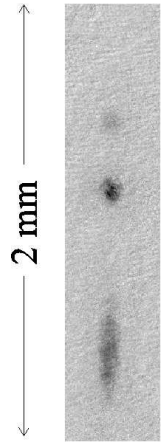}
\end{displaymath}
\caption{Absorption imaging of the multicomponent condensate produced
by a rf pulse. The magnetic trap was on for 1.5~ms after the rf pulse
and the picture was taken after 20~ms of free expansion.}
\label{expell}
\end{figure}

The $m_F=-1,-2$ sublevels are also populated by rf induced multistep
transitions. However, these condensates are quickly expelled by the
magnetic potential and the effect can be observed for shorter times
after the rf pulse. This is evident in the image in Fig.\ref{expell} which is
taken under the same conditions of Fig.\ref{pulse} but with a shorter time
($\Delta$t=1.5~ms) in the magnetic trap. As expected, in addition to the
free--falling $m_F=0$ condensate atoms coupled--out simply by gravity, an
elongated cloud appears, corresponding to atoms in the high--field
seeking states that are repelled from the
trap.

\section{cw atom laser}

Continuously coupling atoms out of a Bose condensate with resonant rf
radiation was first proposed by W.~Ketterle et al. \cite{1ketterle}.
In their paper on the rf output coupler they discuss this scheme and
point out the necessity to have a very stable magnetic field. I.~Bloch
et al. \cite{2bloch} realized a cw atom laser based on rf output
coupling using an apparatus with a very well controlled magnetic
field. They placed a $\mu$--metal shield around the cell where the
condensate forms, achieving residual fluctuations below $10^{-8}$~T.

\begin{figure}[h,t,b]
\begin{displaymath}
\epsfxsize=3cm
\epsfbox{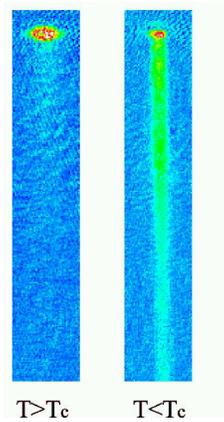}
\end{displaymath}
\caption{Absorption imaging of the atom laser obtained applying a rf
  field for 10~ms and illuminating the atoms 5~ms after switching off
  the
  trap.}
\label{alas}
\end{figure}

We explored the regime of continuous coupling by leaving the rf field
on for at least 10~ms. In this case we observed a stream of atoms
escaping from the trap (Fig.\ref{alas}). The experimental
configuration is similar to the one described in \cite{2bloch}, except
for the fact that our apparatus is not optimized to minimize magnetic
field fluctuations, that are at the level of $10^{-6}$~T.
Nevertheless, our observation demonstrate that these fluctuations do
not prevent the operation of a cw atom laser.

Fig.\ref{alas} shows an absorption image taken after an rf pulse 10~ms
long with an amplitude B$_{rf}$=0.36~$\mu$T. The first picture
corresponds to the temperature of the rubidium atoms being above the
critical temperature (T$_c$) for condensation. In this case a very
weak tail of atoms escaping from the magnetically trapped cloud is
observed. Decreasing the temperature below T$_c$ (second picture of
Fig.\ref{alas}) the beam of atoms leaving the trap becomes sharper and
more collimated.

We have demonstrated a cw atom laser. However, the fluctuations in our
magnetic field strongly influences not only the reproducibility but
also the quality of the extracted beam. The high magnetic field
stability achieved by the M\"{u}nich group allowed the measurement of
the spatial coherence of a trapped Bose gas \cite{16,17} by observing
the interference pattern of two matter waves out--coupled from the
condensate using a rf field composed of two frequencies.

\section{conclusions}

We have illustrated the rich phenomenology arising from the
interaction of an rf field with a $^{87}$Rb condensate originally in
the $F=2$, $m_F=2$ state. We have shown that condensates can be
produced in each of the five Zeeman sublevels and that the relative
populations can be controlled by varying the duration and amplitude of
the rf pulse. We investigated the behaviour of both trapped and
untrapped condensates as a function of the time in the magnetic trap.
At short times we recorded the different behaviour between the atoms
output--coupled under gravity only ($m_F=0$), and those with an
additional impulse due to the magnetic field ($m_F=-1,-2$). We have
also produced a cw atom laser by simply increasing the time duration
of the rf pulse.  In our apparatus no particular care is devoted to
the shielding of unwanted magnetic field. The stability and
homogeneity requirements seem to be less stringent than those
predicted in the pioneering work of \cite{1ketterle} and of those
of the magnetic field implemented by the original cw atom laser
apparatus \cite{2bloch}. This could make the cw atom laser based on rf
out--coupling more generally accessible. Most of the observed
phenomena may be understood using a simple theoretical model; a more
detailed and complete description of the multicomponent condensate
should take into account also the mean field potential and interaction
between different condensates.

Future applications of the experimental set--up we are currently
operating can be foreseen, for instance for the study of collective
excitations induced by the sudden change in the atom number. The
interaction between condensates in different internal states may
possibly be investigated as well as time--domain matter--wave
interferometers using a sequence of rf
pulses.

\begin{acknowledgments}

  We would like to thank M. Prevedelli for his contribution in setting
  up the experiment. This work was supported by the INFM "Progetto di
  Ricerca Avanzata" and by the CNR "Progetto Integrato". We would like
  to thank also D. Lau for careful reading of the
  manuscript.
\end{acknowledgments}

% Bibliography made with BibTeX:

%% apalike is preferred if you have used \kluwerbib, above.

%% Otherwise you may use any .bst style your editor approves.

%\bibliographystyle{apalike}

%\chapbblname{<name of .bbl file>}

%\chapbibliography{<name of .bib file>}

\end{document}